\documentclass[preprint,12pt,authoryear]{elsarticle}

\usepackage{kurz}
\usepackage{amssymb}
\usepackage{mathrsfs}
\usepackage{multirow, hhline}    
\usepackage[table,xcdraw]{xcolor}
\usepackage{amsmath}
\allowdisplaybreaks[4]
\usepackage{caption}
\usepackage{subcaption}
\usepackage{a4wide}
\usepackage{hyperref}
\hypersetup{hidelinks,
	colorlinks=true,
	allcolors=black,
	pdfstartview=Fit,
	breaklinks=true}
\usepackage{enumitem}
\usepackage{algpseudocode}
\usepackage{cleveref}

\usepackage{lineno}

\journal{arXiv}

\begin{document}
	
\begin{frontmatter}



\title{Bridging Simulation and Experiment: A Self-Supervised Domain Adaptation Framework for Concrete Damage Classification}


\author[ISM-BO]{Chen Xu}
\author[ISM-BO]{Giao Vu}
\author[ISM-BO]{Ba Trung Cao\corref{cor}\fnref{fncor}}
\author[ISM-BO,HPC]{Zhen Liu}
\author[TUM]{Fabian Diewald}
\author[HPC]{Yong Yuan}
\author[ISM-BO]{Günther Meschke}

\cortext[cor]{Corresponding author}
\fntext[fncor]{email: ba.cao@rub.de}

\address[ISM-BO]{Institute for Structural Mechanics, Ruhr University Bochum, Universitätsstraße 150, 44801 Bochum, Germany}
\address[HPC]{Department of Geotechnical Engineering, College of Civil Engineering, Tongji University, 1239 Siping Road, 20092 Shanghai, China}
\address[TUM]{Centre for Building Materials, Technical University of Munich, 81245 Munich, Germany}

\begin{abstract}
Reliable assessment of concrete degradation is critical for ensuring structural safety and longevity of engineering structures. This study proposes a self-supervised domain adaptation framework for robust concrete damage classification using coda wave signals. To support this framework, an advanced virtual testing platform is developed, combining multiscale modeling of concrete degradation with ultrasonic wave propagation simulations. This setup enables the generation of large-scale labeled synthetic data under controlled conditions, reducing the dependency on costly and time-consuming experimental labeling. However, neural networks trained solely on synthetic data often suffer from degraded performance when applied to experimental data due to domain shifts.
To bridge this domain gap, the proposed framework integrates domain adversarial training, minimum class confusion loss, and the Bootstrap Your Own Latent (BYOL) strategy. These components work jointly to facilitate effective knowledge transfer from the labeled simulation domain to the unlabeled experimental domain, achieving accurate and reliable damage classification in concrete.
Extensive experiments demonstrate that the proposed method achieves notable performance improvements, reaching an accuracy of 0.7762 and a macro F1 score of 0.7713, outperforming both the plain 1D CNN baseline and six representative domain adaptation techniques. Moreover, the method exhibits high robustness across training runs and introduces only minimal additional computational cost. These findings highlight the practical potential of the proposed simulation-driven and label-efficient framework for real-world applications in structural health monitoring.
\end{abstract}


\end{frontmatter}




\section{Introduction}
\label{sec:Intro}
In modern civil engineering, concrete structures are fundamental to a wide range of infrastructure projects due to their superior strength, durability, and cost efficiency. Nevertheless, concrete inevitably undergoes degradation over time, threatening the structural reliability and potentially leading to catastrophic failures. Therefore, accurate and efficient structural health monitoring (SHM) of concrete structures remains a critical engineering challenge. Under external loading, degradation of concrete typically initiates with the nucleation of microcracks around aggregates, pores, and inherent defects. As the load increases, these microcracks propagate and coalesce into localized macrocracks, ultimately reducing the load-bearing capacity and resulting in structural failure. Due to their significantly smaller scale compared to aggregates, early-stage microcracks are difficult to detect using conventional SHM techniques.

Ultrasonic-based monitoring techniques are particularly attractive for SHM due to their non-invasive, highly sensitive, and cost-effective nature. Since wave velocity is directly influenced by the elastic properties and microstructure of concrete, ultrasonic measurements can provide insight into the internal state of structural materials.
Among these techniques, Coda Wave Interferometry (CWI) \citep{Snieder_2002_science, Snieder_2006_CWI} has shown great promise. CWI works by comparing wave signals measured before and after perturbations caused by microstructural changes. It has successfully identified subtle damage evolution in a variety of materials under varying external conditions \citep{Planes_2013_CWIreview, Wang_2021_CWIbridge, Grabke_2021_CWIconcrete, Grabke_2022_FEMCWI, Shan_2022_metaCWI, Qu_2023_imaging, Liu_2024_rockCWI, Yi_2024_NCWI, Straeter_2023_detection}. With its high sensitivity to slight changes in diffusive media, CWI is able to detect early-stage microcrack initiation and growth, typically beyond the resolution of traditional SHM methods, thus providing a promising approach for early damage detection in concrete structures. 

However, the meaningful interpretation of wave features derived from CWI requires in-depth domain knowledge of concrete behavior, as the relationship between signal variations and mechanical state is complex and often non-unique. Such analyses are typically conducted under well-controlled laboratory conditions. Even then, obtaining error-free, labeled datasets that link ultrasonic signals with precise mechanical states remains a challenge. Conversely, numerical simulations offer a more cost-effective and noise-free alternative, but may lack full physical fidelity. These limitations from both experimental and simulation perspectives present a significant challenge for developing robust, generalizable wave-based damage assessment frameworks applicable to real-world scenarios.

Artificial intelligence (AI) technologies are advancing at an incredible pace, reshaping numerous engineering fields. Among recent developments, physics-informed machine learning (PIML) has emerged as a dynamic interdisciplinary field that integrates prior physical knowledge with machine learning algorithms to tackle complex engineering problems \citep{Raissi_2019_PINN, Karniadakis_2021_PIML, LuLu_2021_DeepONet, Xu2023, Xu2024}. 

Inspired by this paradigm, our research aims to leverage simulated data to train classifiers capable of identifying damage states in concrete structures from coda wave signals. To this end, our previous work \citep{Giao_2021_numerical, Vu_et_al:25} employed physics-based numerical simulations to model the complete damage evolution process in concrete. These simulations generated large volumes of synthetic coda wave data, which were then used to train neural networks for damage state classification. The trained models can be subsequently applied to experimental data collected under real-world conditions.

One of the key advantages of this simulation-driven strategy lies in its ability to considerably reduce the cost and effort associated with collecting labeled experimental data. In many SHM applications, manual labeling is expensive, time-consuming, or even infeasible due to destructive testing requirements and limited access to internal structural states \citep{Liu_2024_crosstl, Wang_2025_disprecon}. By utilizing labeled simulated data, this approach enables concrete damage classification without requiring direct annotation of experimental data, thereby offering a scalable and label-efficient solution for real-world application.

Nevertheless, when neural networks, which have been trained exclusively on simulated data, were tested with real experimental data, a noticeable drop in classification performance was observed. This discrepancy arises primarily from the inevitable simplifications inherent in numerical simulations, which fail to fully replicate the complex cracking behavior of actual concrete structures. As a result, a significant distribution shift \citep{Pan_2010_transfersurvey, Li_2024_crossdomain} exists between the simulated (source domain) and experimental (target domain) data, which severely limits the model’s ability to generalize across domains.

Domain adaptation (DA), as an important subfield of transfer learning, serves as an effective framework for addressing the distribution shift between datasets \citep{Liu_2022_DAreview}. In recent years, several studies have successfully employed DA techniques to transfer knowledge from simulated to experimental domains across various application areas, with machinery fault diagnosis being among the frequently explored topics \citep{Pang_2024_faultvib, Jiang_2024_faultgearbox,Jiang_2025_towards, Huang_2025_intradomain, Zhu_2025_digitalentropy, Xiao_2025_DGsurvey}. Meanwhile, in the field of SHM for civil engineering, Wang et al. \citep{Wang_2022_knowledge} introduced a re-weighted adversarial method to transfer knowledge from finite element models to experimental structures. Giglioni et al. \citep{Giglioni_2024_da} implement Joint Domain Adaptation and Normal Condition Alignment to align features across real bridges and simulations. Zhang et al. \citep{Zhang_2025_dmgtransmiss} developed a structural damage detection method based on the joint maximum discrepancy and adversarial discriminative domain adaptation (JMDAD), enabling transfer learning between numerical models and physical structures. Talaei et al. \citep{Talaei_2025_hybrid} combine Maximum Mean Discrepancy and adversarial DA techniques to reduce domain shift and accurately estimate prestress in concrete bridges under moving loads.

Although DA has shown promise, its application to concrete damage classification, particularly with coda wave data, remains at a preliminary stage and requires further investigation \citep{Weng_2023_unsupervised}. Moreover, existing studies have not fully exploited the latent representations embedded in the target domain to improve accuracy. In our case, the absence of labeled target domain data during training naturally places the problem within the scope of self-supervised learning (SSL) \citep{Liu_2023_sslsurvey, Gui_2024_sslsurvey}. Among various SSL techniques, contrastive learning has demonstrated strong capabilities in extracting more discriminative feature representations from unlabeled datasets, thereby enhancing performance in downstream tasks \citep{Gu_2024_cost, Hoxha_2025_contrast}. Given the lack of labeled experimental data and the need for robust feature representation learning, contrastive learning is well suited to be incorporated as a component of our proposed approach.

In this paper, a domain adaptation framework enhanced with self-supervised learning is proposed to enable reliable knowledge transfer from the simulation domain to the experimental domain. Domain adaptation in our framework is achieved through a combination of Domain-Adversarial Neural Network (DANN) \citep{Ganin_2016_DANN, Li_2023_dmgdetect} and Minimum Class Confusion (MCC) loss \citep{Jin_2020_MCC}. DANN promotes domain-invariant feature learning through adversarial training between a domain discriminator and the feature extractor. Meanwhile, the MCC regularization explicitly reduces inter-class confusion within the target domain, promoting more discriminative and well-separated feature representations. In addition, the Bootstrap Your Own Latent (BYOL) method \citep{Grill_2020_BYOL}, an SSL approach, is incorporated to encourage the network to learn meaningful feature representations from unlabeled experimental data. Our approach requires no labeled experimental data and introduces only minimal additional computational overhead, yet substantially improves the performance of concrete damage classification when applied to real-world experimental scenarios. 

The structure of this paper is as follows: Section~\ref{sec:NUMSIM} summarizes the computational simulations used to generate synthetic data and \Cref{sec:Method} details the proposed domain adaptation framework with self-supervised learning. 
Section~\ref{sec:Datapre} introduces the generation of both simulated data and experimental data for concrete damage classification, and Section~\ref{sec:Result} demonstrates the performance of the proposed method in predicting concrete damage in real-world scenarios. Finally, Section~\ref{sec:Conc} summarizes the main findings, discusses the limitations, and suggests directions for future research.

\section{A Simulation-based Platform for Damage Assessment in Concrete}
\label{sec:NUMSIM}
\begin{figure}[!t]
	\centering
	\includegraphics[width=\textwidth]{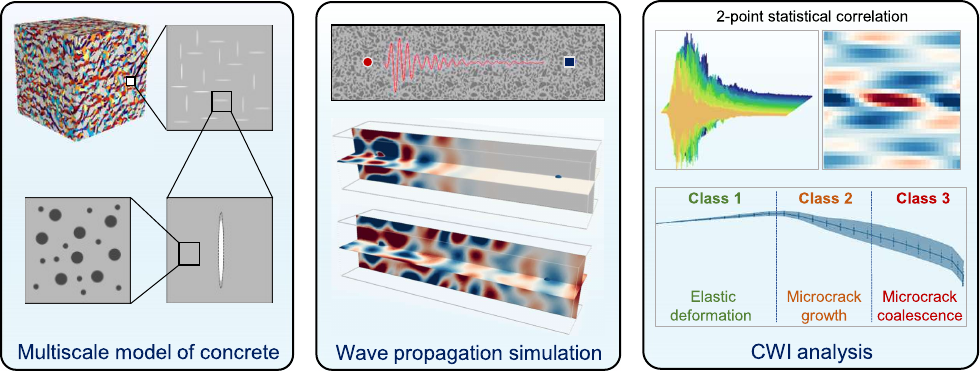}
	\caption{Workflow of the virtual testing platform for damage assessment in concrete structures.}
	\label{fig:sim_scheme}
\end{figure}

To generate synthetic data for training, a simulation-based virtual testing platform developed in \citep{Vu_et_al:25} is employed. 
The process begins with the creation of a realistic concrete specimen using an in-house Concrete Mesostructure Generator. 
The mechanical response of the specimen under uniaxial compression is simulated using a multiscale reduced-order model. At the microscale, microcrack initiation and propagation are modeled within the framework of Linear Elastic Fracture Mechanics (LEFM). At the mesoscale, the corresponding stress and strain fields are computed by solving the integral form of the Lippmann–Schwinger equations, as detailed in \citep{Vu_et_al:21, Vu_et_al:25}.
These deformation snapshots are then used as input for finite difference simulations of ultrasonic wave propagation under multiple source-receiver configurations \citep{Saenger_et_al:05, Saenger_20_rsg}.
A CWI-based analysis is performed to evaluate wavefield changes in response to mechanical loading, enabling comparison with experimental trends and providing labeled damage states for supervised learning.
Finally, spatially informative patterns are captured from the simulated ultrasonic signals using two-point statistical correlations, forming the input for classification.
The overall computational workflow is illustrated in \Cref{fig:sim_scheme}.

\subsection{Multiscale and Ultrasonic Wave Propagation Simulation in Concrete}

\paragraph{Multiscale modeling}
Multiscale modeling offers a rigorous, physics-based approach to linking the macroscopic material behavior in concrete with its underlying microstructural evolution during loading, without the need for empirical assumptions \citep{Matouvs_et_al:17}. 
It operates through two fundamental mechanisms: homogenization, which transfers microscale mechanical properties (e.g., stiffness and stress) to the macroscale, and localization, which maps macroscopic strain fields back to the microstructure. 
This bidirectional coupling enables accurate simulation of complex damage phenomena across scales.

Homogenization, which is typically computationally intensive in multiscale modeling, is addressed using a k-means clustering-based model reduction technique. This method groups similar material points and assigns each cluster a representative subproblem, thereby improving computational efficiency \citep{Liu_2016_SCcluster}. 
The concrete mesostructure is discretized using a voxel-based representation, in which aggregates and mortar are explicitly resolved. 
At the microscale, the mortar matrix contains three orthogonal families of penny-shaped microcracks, each characterized by specific orientation, volume fraction, radius, and thickness. 
These microcracks serve as the principal mechanism for simulating diffuse damage under compressive loading.

Localization is performed by solving the Lippmann–Schwinger equation to compute mesoscopic strain fields, which in turn drive microcrack evolution based on Linear Elastic Fracture Mechanics. 
Homogenization then updates the macroscopic stiffness tensor using a modified Interaction Direct Derivative with evolving microcrack distribution \citep{Vu_et_al:21}. 
Despite using a simplified crack representation, the model effectively captures load-induced anisotropy and early-stage damage in concrete.

\paragraph{Wave propagation modeling}
Wave propagation is modeled by solving elastodynamic wave equations using finite difference schemes on Rotated Staggered Grids \citep{Saenger_20_rsg}.
To account for the decaying characteristics of ultrasonic wave fields traversing through concrete, a viscoelastic material model with two Maxwell bodies in employed \citep{Saenger_et_al:05}.
The ultrasonic wave source is modeled as a six-component tensor to accurately replicate the radiation pattern of real transducers. Meanwhile, the recorded signals are the temporal stresses over a prescribed time interval.

\subsection{Coda Wave Interferometry (CWI) for Generating Training Outputs}

Relative velocity variation ($dv/v$) and cross-correlation (CC) are key metrics for quantifying time shifts and waveform similarity in ultrasonic and seismic signals. 
The $dv/v$ metric captures subtle, uniform changes in wave speed linked to microstructural changes, such as crack closure, compaction, or microcracking, while a decrease in the CC signal is well suited to indicate heterogeneous changes of the microstructure, such as localized cracking or stress-induced anisotropy. 
In this study, $dv/v$ is computed using the stretching technique, which aligns perturbed and reference signals by adjusting a stretch factor $\alpha$ \citep{niederleithinger2018processing}. The value of $\alpha$ that maximizes the CC between the two signals is taken as the $dv/v$.
The $dv/v$ metric serves as a reliable indicator for detecting and classifying internal damage in concrete and is calculated as:
\begin{equation}
\frac{dv}{v} = \arg\max_{\alpha \in \mathbb{R}} \, \text{CC}(\alpha) = \arg\max_{\alpha \in \mathbb{R}} \frac{\int_{t_1}^{t_2} u_u(t) u_p(t(1 + \alpha)) \, dt}{\sqrt{\int_{t_1}^{t_2} u_u^2(t) \, dt \int_{t_1}^{t_2} u_p^2(t(1 + \alpha)) \, dt}} \times 100\%.
\end{equation}

As a preliminary analysis, CWI is employed to study the influence of material degradation on ultrasonic wave velocity variations. By comparing numerical and experimental results, it serves as a practical tool to validate the mechanical–ultrasonic coupling implemented in the simulations. However, while the obtained relative velocity changes provide good indication regarding the intensity of material degradation, the state of degradation cannot be directly derived.

\subsection{Two-Point Statistical Correlation for Generating Training Inputs}
Traditional CWI features, such as $dv/v$ and CC, effectively capture global variations in signals but provide limited insight into specific damage mechanisms. 
Two-point statistics, however, yield richer and more detailed information on the type and severity of damage by capturing spatial correlations within the ultrasonic waveforms \citep{cecen2016versatile,brough2017materials}.
Their high dimensionality also enhances sensitivity and robustness to noise,  making them effective for machine learning-based damage classification tasks.
To compute two-point statistics from 1D ultrasonic signals, consider two normalized ultrasonic waveforms: a reference signal $\tilde{u}_u[t]$ and a perturbed signal $\tilde{u}_p[t]$. 
The two-point correlation function is obtained by correlating these signals across discrete temporal shifts. Specifically, the statistics is calculated as follows:
%
\begin{equation}
f[\chi] = \frac{1}{\delta[\chi]} \sum_{t} \tilde{u}_u[t] \tilde{u}_p[t+\chi],
\end{equation}
where $\chi$ denotes the discrete time lag in the one-dimensional temporal domain, and the normalization factor $\delta[\tau]$ accounts for variations in the number of overlapping samples at each lag. By directly utilizing this correlation approach, the resulting two-point statistics effectively encode detailed temporal features, facilitating accurate detection and classification of subtle structural changes.

\section{The Proposed Self-Supervised Domain Adaptation Framework}
\label{sec:Method}
\begin{figure}[!t]
	\centering
	\includegraphics[scale=0.9]{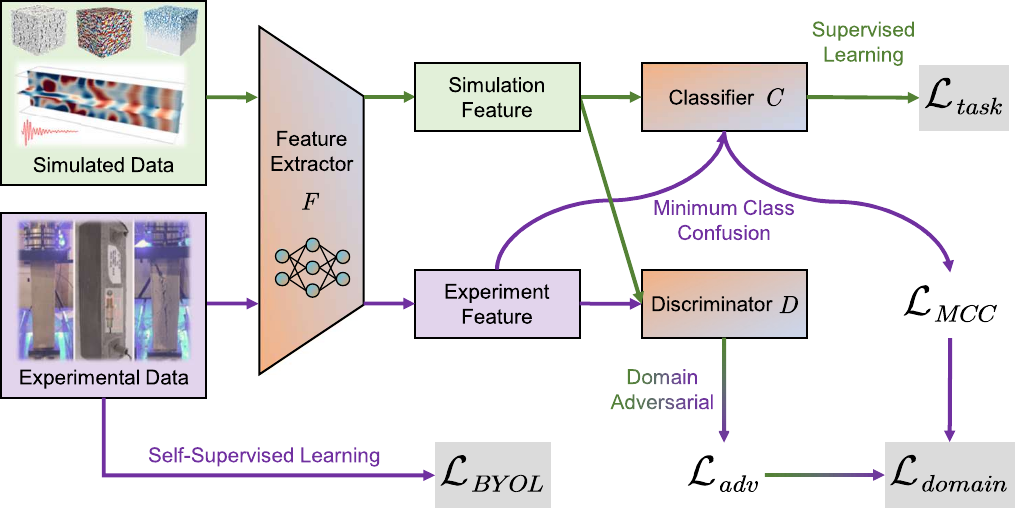}
	\caption{Schematic of the proposed self-supervised domain adaptation framework for concrete damage classification using coda wave signals. The architecture integrates supervised learning on simulated data with domain adversarial training, minimum class confusion (MCC), and self-supervised BYOL strategy on experimental data. These components work synergistically to align feature distributions and improve generalization across domains.}
	\label{fig:method}
\end{figure}

The fundamental concepts behind our proposed framework are presented in detail in the following subsections. A schematic of the entire framework is illustrated in Fig.~\ref{fig:method}, highlighting each major component and the data flow between them. 
The proposed framework consists of a shared feature extractor $F$, a classifier $C$, and a domain discriminator $D$, jointly trained using both simulated and experimental data. The classifier $C$ learns under supervision from the labeled simulation data, while the discriminator $D$ encourages domain-invariant feature learning through adversarial training and the MCC objective. In parallel, a BYOL branch introduces self-supervised learning on the experimental data to improve feature representations without requiring labels. All modules are concurrently optimized through three loss components: $\mathcal{L}_{task}$, $\mathcal{L}_{domain}$, and $\mathcal{L}_{BYOL}$, enabling robust knowledge transfer from simulation to experiment.

\subsection{1D Convolutional Neural Network (CNN) for Classification}
\label{sec:cnn}
Let us first focus on the most fundamental component for the classification task: a 1D CNN. With its extraordinary performance, the 1D CNN has become the \textit{de facto} standard architecture for signal processing across various research areas \citep{Kiranyaz_2021_1dcnnsurvey, Li_2022_cnnsurvey, Xiong_2025_reacgan}.

\begin{figure}[!t]
	\centering
	\includegraphics[scale=0.9]{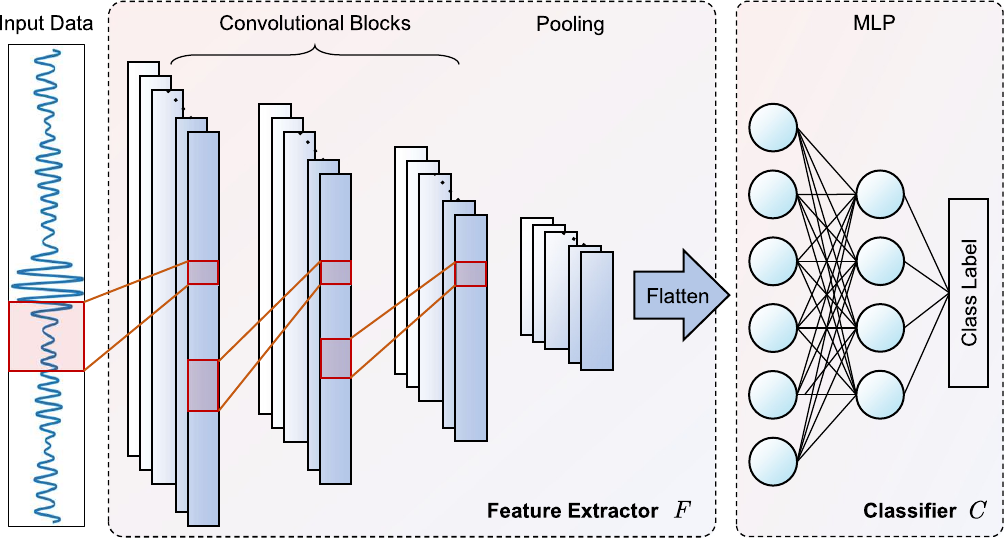}
	\caption{Architecture of a 1D CNN model consisting of a convolutional feature extractor $F$ and an MLP classifier $C$.}
	\label{fig:1dcnn}
\end{figure}

\begin{figure}[!t]
	\centering
	\includegraphics[scale=0.92]{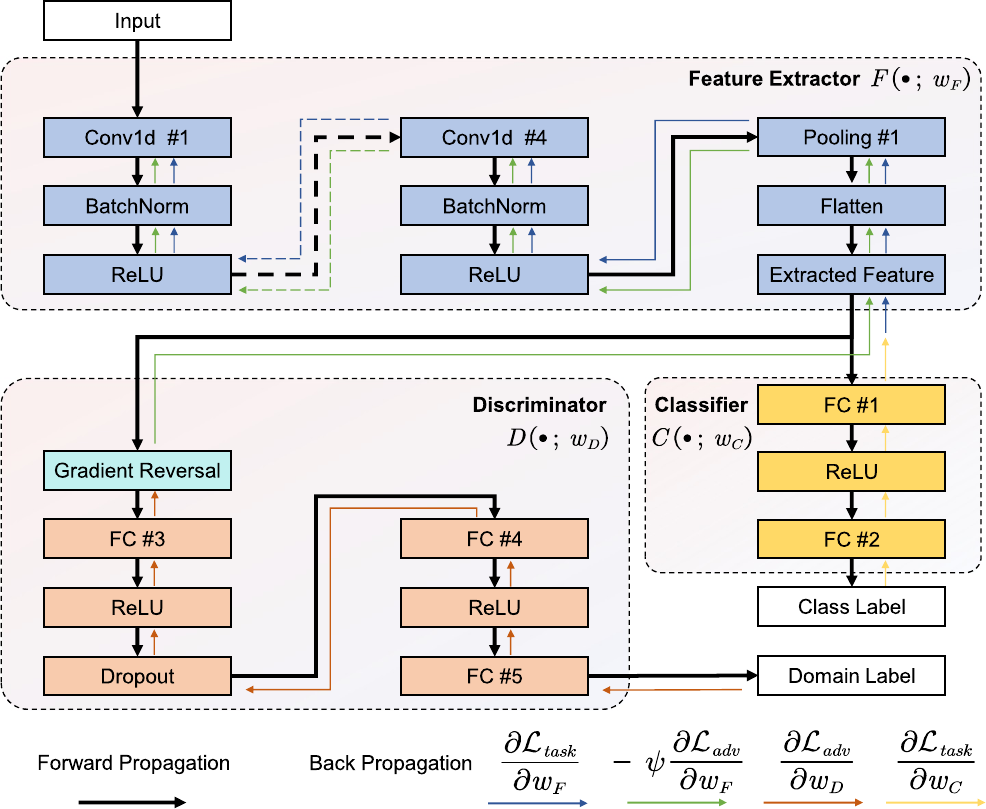}
	\caption{Detailed architecture of the domain adversarial training module within the proposed framework, including a feature extractor $F(\cdot; w_F)$, a classifier $C(\cdot; w_C)$, and a domain discriminator $D(\cdot; w_D)$. Forward propagation paths are shown in black, while colored arrows indicate component-specific gradient flows during backpropagation. Detailed layer-wise configurations for convolutional (Conv1d) and fully connected (FC) layers are provided in Table~\ref{tab:nnparams}.}
	\label{fig:dann}
\end{figure}

As illustrated in Fig.~\ref{fig:1dcnn}, a 1D CNN can be viewed as a combination of a feature extractor $F$ and a classifier $C$. The feature extractor $F$ is designed to extract feature representations from the input data and comprises several stacked convolutional blocks. In each block (shown in Fig.~\ref{fig:dann}), the 1D convolutional layers (Conv1d) can extract local features by sliding convolutional kernels over the input and capture spatial information through parameter sharing. Each convolutional layer is followed by a batch normalization layer and a rectified linear unit (ReLU) activation function. Subsequently, a pooling layer is employed to reduce the dimensionality of the data, and the output is then flattened to obtain the extracted features. The classifier $C$, implemented as a multi-layer perceptron (MLP), finally processes the features from $F$ and assigns them to appropriate classes.

The dataset obtained from numerical simulation is denoted as $\mathcal{D}_s=\left\{ x_{i}^{s}, y_{i}^{s} \right\}_{i}^{N_s}$, with its underlying data distribution represented by $p_s$, where the subscript $s$ stands for source domain. A 1D CNN is trained on $\mathcal{D}_s$ in a fully supervised manner. The input samples $x^s$ are first processed by the feature extractor $F$, and the resulting features are fed into the classifier $C$ to produce the predicted labels $C\left( F\left( x^s \right) \right)$. The model is optimized by minimizing the cross-entropy loss $\ell$ between the predicted labels and the corresponding ground-truth labels $y^s$. The supervised classification loss is therefore formulated as follows:
\begin{equation}
\begin{aligned}
\mathcal{L} _{task}\left( F,C \right) =\mathbb{E} _{\left( x,y \right) \thicksim p_s}\left[ \ell \left( C\left( F\left( x \right) \right) , y \right) \right].
\end{aligned}
\label{eq:cnnloss}
\end{equation}

\subsection{Adversarial Domain Adaptation with Minimum Class Confusion (MCC)}
\label{sec:dann}
The dataset obtained from experiment is denoted as $\mathcal{D}_t=\left\{ x_{i}^{t}, y_{i}^{t} \right\}_{i}^{N_t}$, with its corresponding data distribution being $p_t$, where the subscript $t$ stands for target domain. When there is a discrepancy between the data distributions of the source and target domains ($p_s\ne p_t$), also referred to as domain shift \citep{Pan_2010_transfersurvey}, directly applying the model trained on the source domain (i.e. the simulation-based 1D CNN described in Section~\ref{sec:cnn}) leads to a significant performance degradation on the target domain (experimental data). To tackle this problem, the DANN proposed in \citep{Ganin_2016_DANN} is adopted to achieve domain adaptation. This approach forces the model to learn features that are robust to domain shifts, thereby enhancing its generalization capability on the target domain. As shown in Fig.~\ref{fig:dann}, a third component is introduced, namely the domain discriminator $D$, which establishes an adversarial training process between the feature extractor $F$ and the discriminator $D$. The discriminator takes the extracted features as input and outputs domain labels, where the labels for the source and target domains are set to 0 and 1, respectively. The loss function of DANNs can be written as:
\begin{equation}
\begin{aligned}
\mathcal{L} _{adv}\left( F,D \right) =\mathbb{E} _{x\thicksim p_s}\left[ \ell \left( D\left( F\left( x \right) \right) , 0 \right) \right] +\mathbb{E} _{x\thicksim p_t}\left[ \ell \left( D\left( F\left( x \right) \right) , 1 \right) \right].
\end{aligned}
\label{eq:advloss}
\end{equation}

The underlying idea behind Eq.~\eqref{eq:advloss} is that the domain discriminator $D$ continuously improves its ability to distinguish between features from the source and target domains by minimizing $\mathcal{L} _{adv}$, while the feature extractor $F$ aims to learn features that are hard for the discriminator $D$ to differentiate by maximizing $\mathcal{L} _{adv}$. This adversarial setup gives rise to a minimax game between two players $F$ and $D$, which compels the model to extract domain-invariant features that perform well across both the source and target domains. To achieve this, as shown in Fig.~\ref{fig:dann}, a Gradient Reversal Layer (GRL) is inserted at the beginning of the domain discriminator $D$. The GRL serves as an identity function during forward propagation, but during backpropagation it multiplies the gradient by a negative constant $-\psi $ (where $\psi>0$ is a hyperparameter controlling the adversarial strength), that is:
\begin{equation}
\begin{aligned}
R_{\psi}\left( k \right) =k, \quad \frac{\partial R_{\psi}\left( k \right)}{\partial k}=-\psi I.
\end{aligned}
\end{equation}

\begin{figure}[!t]
	\centering
	\includegraphics[scale=0.9]{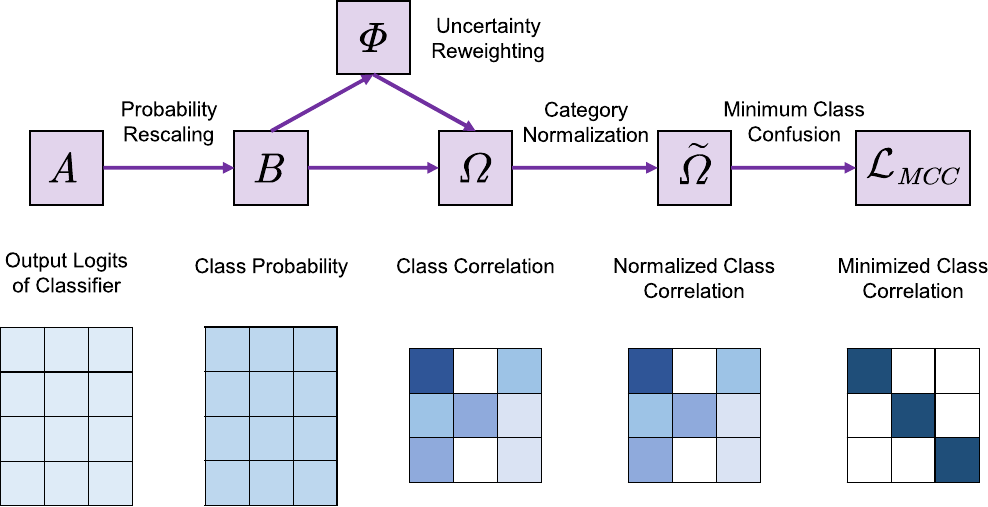}
	\caption{Overview of the Minimum Class Confusion (MCC) loss computation.}
	\label{fig:mcc}
\end{figure}

In order to alleviate class ambiguity in the target domain, the Minimum Class Confusion (MCC) loss \citep{Jin_2020_MCC} is introduced as a complementary component to the domain adaptation process. Unlike conventional feature alignment strategies that operate in the latent feature space, MCC directly works on the output logits of classifier. It strives to reduce pairwise confusion among classes by leveraging the output distribution, thus encouraging clearer decision boundaries in the target domain. The whole procedure for computing the MCC loss is illustrated in Fig.~\ref{fig:mcc}.

Let $\mathbf{A} \in \mathbb{R}^{N_b \times |\mathcal{C}|}$ denote the classifier logits for a batch of $N_b$ unlabeled target domain samples, where $|\mathcal{C}|$ is the number of classes. To prevent overconfident predictions from dominating the confusion estimation, a temperature scaling technique is applied to calibrate the logits:
\begin{equation}
\begin{aligned}
\hat{B}_{ij}=\frac{\exp \left( A_{ij}/T \right)}{\sum_{j\prime =1}^{|\mathcal{C} |}{\exp \left( A_{ij\prime}/T \right)}},
\end{aligned}
\end{equation}
where $T$ is a temperature hyperparameter controlling the smoothness of the prediction probabilities, and $\hat{B}_{ij}$ represents the probability that the $i$-th sample in the batch belongs to class $j$. 

Following the probability rescaling step, MCC constructs a preliminary class correlation matrix to quantify the degree to which two classes are simultaneously activated in the model’s predictions:
\begin{equation}
\begin{aligned}
\varOmega _{jj^{\prime}}=\hat{\mathbf{b}}_{.j}^{\top}\hat{\mathbf{b}}_{.j^{\prime}},
\end{aligned}
\end{equation}
where $\hat{\mathbf{b}}_{.j} \in \mathbb{R}^{N_b}$ contains the predicted probabilities for class $j$ across the batch. A higher value of $\varOmega_{jj^{\prime}}$ indicates that the model tends to assign high probabilities to both class $j$ and class $j'$ for the same set of samples, which reflects greater ambiguity between these two classes. 

Since not all samples contribute equally to the estimation of class confusion, MCC employs an entropy-based weighting mechanism. The entropy (uncertainty) of the prediction $\hat{\mathbf{b}}_{i.} \in \mathbb{R}^{|\mathcal{C}|}$ of sample $i$ is computed as
\begin{equation}
\begin{aligned}
H(\hat{\mathbf{b}}_{i.})=-\sum_{j=1}^{|\mathcal{C} |}{\hat{B}_{ij}\log \hat{B}_{ij}},
\end{aligned}
\end{equation}
based on this entropy, a weight reflecting the importance of sample $i$ in modeling class confusion is defined as
\begin{equation}
\begin{aligned}
\varPhi _{ii}=\frac{N_b\left( 1+\exp \left( -H(\hat{\mathbf{b}}_{i.}) \right) \right)}{\sum_{i^{\prime} =1}^{N_b}{\left( 1+\exp \left( -H(\hat{\mathbf{b}}_{i.}) \right) \right)}},
\end{aligned}
\end{equation}
with this uncertainty reweighting, the refined class correlation is given by:
\begin{equation}
\begin{aligned}
\varOmega _{jj^{\prime}}=\hat{\mathbf{b}}_{.j}^{\top}\boldsymbol{\varPhi}\hat{\mathbf{b}}_{.j^{\prime}},
\end{aligned}
\end{equation}
where $\boldsymbol{\varPhi}$ is the corresponding diagonal matrix for $\varPhi _{ii}$.

To mitigate class imbalance within each batch, the refined correlation undergoes an additional step of category normalization:
\begin{equation}
\begin{aligned}
\tilde{\varOmega}_{jj^{\prime}}=\frac{\varOmega _{jj^{\prime}}}{\sum_{j^{''}=1}^{|\mathcal{C} |}{\varOmega _{jj^{''}}}},
\end{aligned}
\end{equation}
the resulting $\tilde{\varOmega}_{jj^{\prime}}$ can effectively capture the confusion between class $j$ and $j^{\prime}$. The MCC loss is then computed as the average normalized confusion between all pairs of different classes:
\begin{equation}
\begin{aligned}
\mathcal{L} _{MCC}=\frac{1}{|\mathcal{C} |}\sum_{j=1}^{|\mathcal{C} |}{\sum_{j^{\prime}\ne j}^{|\mathcal{C} |}{\tilde{\varOmega}_{jj^{\prime}}}}.
\end{aligned}
\label{eq:mccloss}
\end{equation}

The final domain adaptation objective combines both the adversarial loss in Eq.~\eqref{eq:advloss} and the MCC loss in Eq.~\eqref{eq:mccloss}, and is expressed as:
\begin{equation}
\begin{aligned}
\mathcal{L} _{domain}=\mathcal{L} _{adv}+\mathcal{L} _{MCC},
\end{aligned}
\end{equation}

\subsection{Self-Supervised Learning: Bootstrap Your Own Latent (BYOL)}
\label{sec:byol}
In order to further extract experimental features that are more meaningful for the downstream classification task, a SSL method known as BYOL \citep{Grill_2020_BYOL} is adopted. Its core idea is to learn useful feature representations by aligning two sets of networks, rather than relying on contrasting negative pairs \citep{He_2020_Moco, Chen_2020_Simclr}. 

\begin{figure}[!t]
\centering
\includegraphics[scale=0.92]{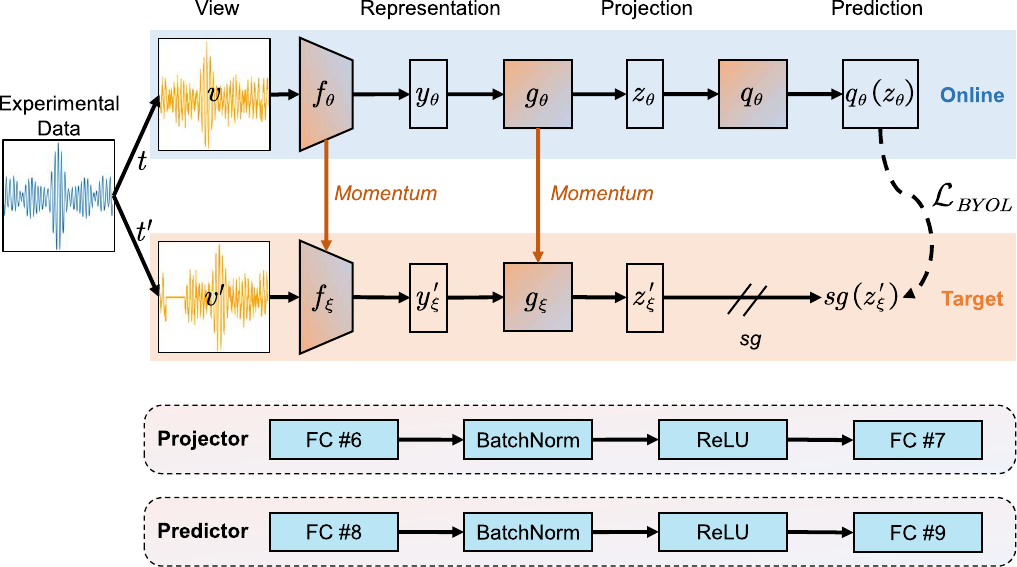}
\caption{Architecture of the BYOL-based self-supervised learning module. The layer configurations of fully connected (FC) layers in the projector and predictor are detailed in Table~\ref{tab:nnparams}.}
\label{fig:byol}
\end{figure}

BYOL consists of an online network and a target network, with their detailed architecture depicted in Fig.~\ref{fig:byol}. The online network is composed of an encoder $f_{\theta}$, a projection head $g_{\theta}$, and a predictor $q_{\theta}$; while the target network only contains an encoder $f_{\xi}$ and a projection head $g_{\xi}$. 
The encoders of both the online and target networks adopt the identical architecture as the feature extractor $F$ described in Sections~\ref{sec:cnn} and \ref{sec:dann}. Notably, the encoder $f_{\theta}$ of the online network shares its network parameters with the feature extractor $F$. The projection heads and predictor are implemented as MLPs.

For each input sample, two different views $v$ and $v'$ are generated through random data augmentations $t$ and $t'$, respectively. The view $v$ is input into the online network where it is first transformed into a primary feature representation $h_{\theta}$ by the encoder $f_{\theta}$, then mapped to $z_{\theta}$ via the projection head $g_{\theta}$, and finally processed by the predictor $q_{\theta}$ to produce the prediction $q_{\theta}(z_{\theta})$. The computation process for $v'$ in the target network follows a similar procedure to that of the online network, except that the predictor is excluded. The objective of BYOL is to make the online network's prediction $q_{\theta}(z_{\theta})$ to approximate the target network's projection $z_{\xi}^{\prime}$. To this end, mean squared error is used as the loss function, which can be expressed as:
\begin{equation}
\begin{aligned}
\mathcal{L} _{BYOL}=\left\| \frac{q_{\theta}\left( z_{\theta} \right)}{\left\| q_{\theta}\left( z_{\theta} \right) \right\| _2}-\frac{z_{\xi}^{\prime}}{\left\| z_{\xi}^{\prime} \right\| _2} \right\| _{2}^{2}=2-2\cdot \frac{\left< q_{\theta}\left( z_{\theta} \right) , z_{\xi}^{\prime} \right>}{\left\| q_{\theta}\left( z_{\theta} \right) \right\| _2\left\| z_{\xi}^{\prime} \right\| _2}.
\end{aligned}
\end{equation}

The parameters $\theta$ of the online network are updated by optimizing the training loss through gradient descent, while the target network parameters $\xi$ are not directly updated. Instead, they are updated using an Exponential Moving Average (EMA) method as follows:
\begin{equation}
\begin{aligned}
\xi \gets \tau \xi +\left( 1-\tau \right) \theta,
\end{aligned}
\end{equation}
where $\tau$ is the momentum coefficient.

\subsection{Model Training and Evaluation}
\label{sec:train}
The proposed method encompasses a total of seven neural network components: the encoder of the online network $f_\theta$ (which is equivalent to the feature extractor $F$), the projection head of the online network $g_\theta$, the predictor of the online network $q_\theta$, the encoder of the target network $f_\xi$, the projection head of the target network $g_\xi$, the classifier $C$, and the domain discriminator $D$. Table~\ref{tab:nnparams} provides detailed parameter settings for all of these components. The overall loss function is given by
\begin{equation}
\begin{aligned}
\mathcal{L}_{total}=\mathcal{L} _{task}+\mathcal{L} _{domain}+\mathcal{L} _{BYOL}.
\end{aligned}
\end{equation}

\begin{table}[!t]
\centering
\caption{Layer-wise configuration of neural networks used in the proposed method.}
\begin{tabular}{c|c}
\hline
Name of Layer     & Setting                                              \\ \hline
Conv1d \#1        & nn.Conv1d(1, 16, kernel\_size=9, stride=2, padding=4)  \\
Conv1d \#2        & nn.Conv1d(16, 32, kernel\_size=7, stride=2, padding=3) \\
Conv1d \#3        & nn.Conv1d(32, 32, kernel\_size=5, stride=2, padding=2) \\
Conv1d \#4        & nn.Conv1d(32, 32, kernel\_size=3, stride=2, padding=1) \\
Pooling \#1       & nn.AdaptiveAvgPool1d(16)                               \\
FC \#1            & nn.Linear(512, 32)                                     \\
FC \#2            & nn.Linear(32, 3)                                       \\
FC \#3 and FC \#6 & nn.Linear(512, 256)                                    \\
FC \#4            & nn.Linear(256, 64)                                     \\
FC \#5            & nn.Linear(64, 1)                                       \\
FC \#7 and FC \#9 & nn.Linear(256, 128)                                    \\
FC \#8            & nn.Linear(128, 256)                                    \\ \hline
\end{tabular}
\label{tab:nnparams}
\end{table}

\begin{table}[!t]
\centering
\caption{Hyperparameters settings in the codes.}
\begin{tabular}{c|c}
\hline
Parameter                  & Value    \\ \hline
Training  epochs           & 500      \\
Batch size                 & 128      \\
Learning rate (except $D$) & 5.00e-05 \\
Learning rate (for $D$)    & 1.00e-05 \\
Weight decay               & 1.00e-04 \\
$\psi $                    & 1.0      \\
$T$                        & 2.5      \\
$\tau$                     & 0.99     \\ \hline
\end{tabular}
\label{tab:hyperparams}
\end{table}

The parameters were optimized using the Adam optimizer, and all hyperparameter details are provided in Table~\ref{tab:hyperparams}. All the input data have to be scaled in the range of (-1, 1) by min-max normalization. The data augmentation employed in BYOL consists of Gaussian noise addition and random shifting. Specifically, Gaussian noise with a magnitude of 0.1 is added to the signal. Each signal is circularly shifted by a random integer value within ±10\% of its total length. All machine learning processes were implemented in Pytorch and trained on an NVIDIA T4 GPU via Google Colab.

The performance of the proposed model is evaluated using two metrics: accuracy and the macro F1 score, with their calculation formulas given below:
\begin{gather}
Accuracy=\frac{TP_1+TP_2+TP_3}{N_s+N_t},
\\
Precision_i=\frac{TP_i}{TP_i+FP_i}, \quad Recall_i=\frac{TP_i}{TP_i+FN_i},
\\
F1_i=2\times \small{\frac{Precision_i\times Recall_i}{Precision_i+Recall_i}}, 
\\
Macro\,\,F1=\small{\left( F1_1+F1_2+F1_3 \right) /3},
\end{gather}
here, the index $i$ indicates the class, $TP_i$ (True Positives) denotes the number of positive samples that are correctly predicted, $TN_i$ (True Negatives) represents the number of negative samples that are correctly predicted, $FP_i$ (False Positives) is the number of negative samples that are incorrectly predicted as positive, and $FN_i$ (False Negatives) is the number of positive samples that are incorrectly predicted as negative.

\section{Data Preparation}
\label{sec:Datapre}
\subsection{Laboratory Tests for Experimental Dataset (Target Domain)}
\label{sec:exp_compresstest}
The experiment data used in this paper is taken from \citep{Diewald_2022_exp}. All experiments were performed using a consistent concrete mix composed of ordinary Portland cement (350 kg/m³, water–cement ratio 0.45) and quartzitic crushed aggregates graded according to an AB8 curve. A polycarboxylatether-based superplasticizer was added to improve workability and reduce paste viscosity, ensuring proper coupling between embedded sensors and the matrix. The Young’s modulus obtained from pre-tests was 84.6 GPa for the quartzitic aggregates and 27.1 GPa for the mortar, while the Poisson’s ratio was 0.12 and 0.19, respectively. Compressive strengths were measured as 368.0 MPa for the aggregates and 80.3 MPa for the mortar. The fresh concrete had an air void content of 3.9\%. Specimens were water-cured at $20\,^\circ\mathrm{C}$ for 7 days, followed by storage at 65\% RH and $20\,^\circ\mathrm{C}$. All experiments were carried out 28 days after casting.

\begin{figure}[!t]
	\centering
	\includegraphics[scale=0.95]{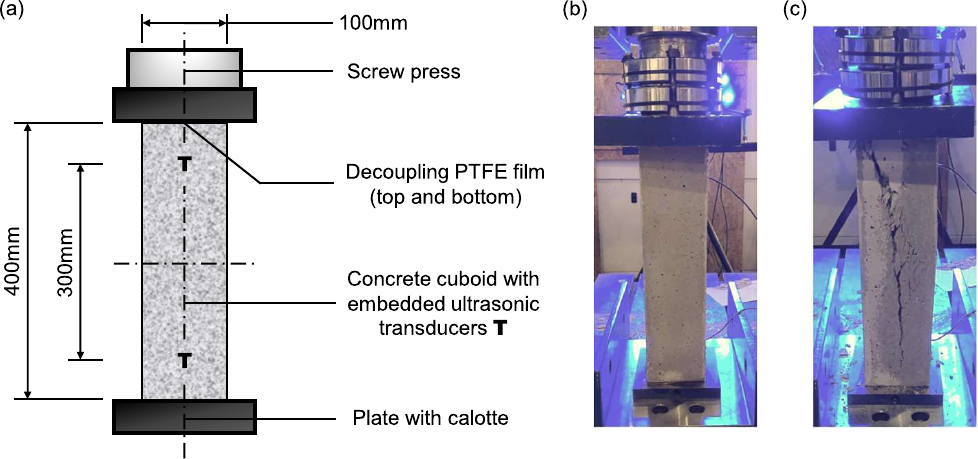}
	\caption{Experimental setup for the uniaxial compression test: (a) schematic of the concrete specimen with dimensions, (b) photo of a specimen during loading, and (c) photo of a specimen after failure.}
	\label{fig:exp_setup}
\end{figure}

As shown in Fig.~\ref{fig:exp_setup}, three cuboid specimens ($100\times100\times400$ mm) were used for the compression tests, each equipped with a pair of transducers symmetrically positioned 300 mm apart. To minimize friction and prevent multiaxial stress near the end faces, PTFE films were placed between the sawn specimen ends and the hydraulic press. The average compressive strength of three identical reference specimens without embedded transducers was measured to be $f_c = 42.9$ MPa. Each test began with a $5\%f_c$ preload to relieve internal stresses, followed by uniaxial compressive loading at 50 N/s. Ultrasonic signals were collected at a rate of 0.1 Hz. Maximum loads of 100\%, 60\%, and 30\% of the compressive strength $f_c$ were applied to the three specimens, respectively, after which each was unloaded at 50 N/s.

\begin{figure}[!t]
	\centering
	\includegraphics[scale=1.0]{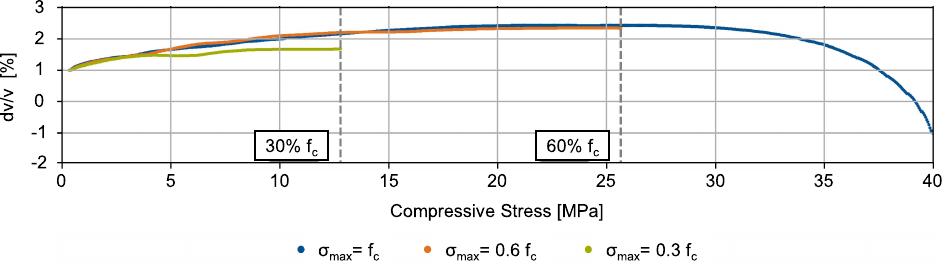}
	\caption{Relative velocity change ($dv/v$) and signal correlation coefficient (CC) for three specimens in laboratory tests, presented with respect to compressive stress during loading up to 100\%, 60\%, and 30\% of the compressive strength $f_c$.}
	\label{fig:exp_result}
\end{figure}

The relative velocity change ($dv/v$) was evaluated using a fixed, non-perturbed reference signal obtained at the unloaded state. Both parameters were analyzed as a function of compressive stress during the uniaxial loading process for three specimens subjected to different maximum loads, as shown in Fig.~\ref{fig:exp_result}. The results show that the $dv/v$ increases with increasing load up to 60\% of $f_c$ (25.6 MPa), after which a decline is observed, indicating the beginning of material damage. For the two non-failed specimens, unloading led to a partial recovery of $dv/v$, reaching 0.6\% and 0.9\% for the 30\% and 60\% $f_c$ cases, respectively, indicating irreversible microstructural changes. Finally, a total of 2133 samples were collected in the experimental dataset $\mathcal{D}_t$, with 753 for Class 1, 979 for Class 2, and 401 for Class 3. 

\subsection{Numerical Modeling for Simulated Dataset (Source Domain)}

\subsubsection{Multiscale Simulation Model}
\label{sec:sim_multiscale_sim}
\begin{figure}[!t]
	\centering
	\includegraphics[scale=1.0]{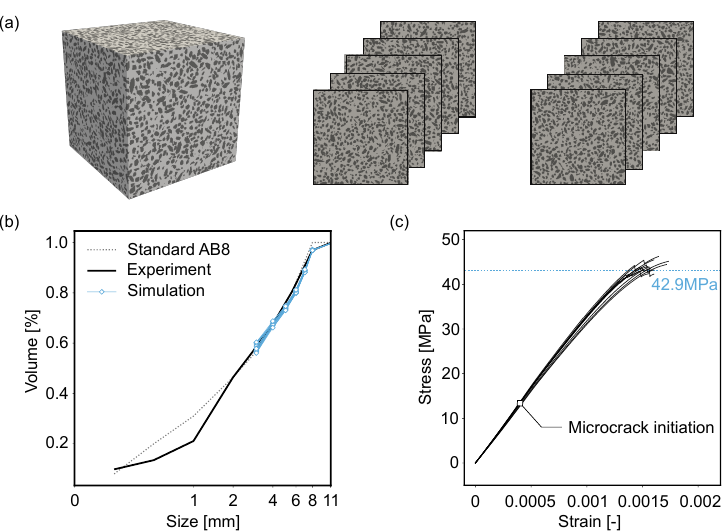}
	\caption{Multiscale model for concrete specimen. (a) A virtual concrete specimen with representative 2D slices; (b) Validation for volume distribution of aggregates; (c) Validation for stress–strain response of concrete specimen under uniaxial compression.}
	\label{fig:sim_multiscale_validation}
\end{figure}

The synthetic data in this paper is generated from the simulation examples presented in \citep{Vu_et_al:25}.
The examples show synthetic samples replicating experimental concrete specimens from \Cref{sec:exp_compresstest}, see \Cref{fig:sim_multiscale_validation}(a). 
High-resolution voxel-based models were constructed using $201^3$ voxels at a resolution of 0.5 mm. 
Aggregates larger than 3 mm were explicitly resolved, achieving a 35\% volume fraction with a standard deviation of 2.26\%, closely matching the experimentally measured particle size distribution. 
Smaller aggregates (volume fraction 35.5\%) were embedded within the hardened cement paste matrix using the Mori-Tanaka homogenization scheme to form the mortar phase.
It can be seen in \Cref{fig:sim_multiscale_validation}(b) that comparison with experimentally measured distribution shows good agreement.
The multiscale model of concrete under compression is summarized as follows. The model assumes that both mortar and aggregate contain three mutually orthogonal microcrack families, each defined by specific orientations, topology, and densities. During the initial stage of compressive loading, microcracks oriented perpendicular to the loading direction tend to close partially, resulting in a slight increase in stiffness. As the load increases to medium and high levels, lateral expansion induces tensile stresses in the transverse directions. This promotes the propagation of microcracks, whose evolution follows the principles of Linear Elastic Fracture Mechanics (LEFM). At the macroscopic scale, this process manifests as progressive stiffness degradation. Further details can be referred to in \cite{Vu_et_al:21,Vu_et_al:25}.

To manage computational complexity, the voxel data were reduced to 72 representative clusters, with 64 for mortar and 8 for aggregates, reducing the model from over 8 million material points to a tractable size without compromising accuracy. 
\Cref{fig:sim_multiscale_validation}(c) illustrates the validation of the multiscale simulation model in a prior study \citep{Vu_et_al:21} for stress-strain response of concrete specimen under uniaxial compression, offers a robust balance between computational efficiency and fidelity.
In total, 10 concrete specimens of size 10 cm with a maximum grain size 8 mm (standard AB8) are generated for the test considering the varying of aggregate inside the concrete specimen.
\subsubsection{Ultrasonic Wave Propagation Simulation}
\begin{figure}[!t]
	\centering
	\includegraphics[scale=1.0]{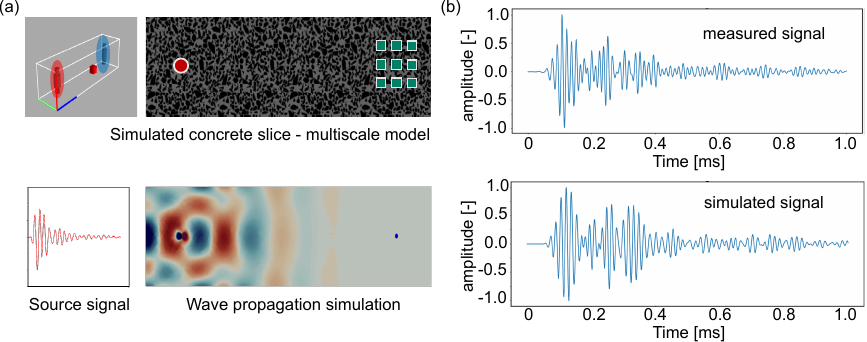}
	\caption{Wave propagation simulation in concrete. (a) Summary of the 3D simulation setup; (b) Validation of simulated wave signals.}
	\label{fig:sim_wave_validation}
\end{figure}

The multiscale model from \Cref{sec:sim_multiscale_sim} is used to simulate the propagation of ultrasonic wave fields through virtual specimens.
The simulation is designed around prismatic concrete specimens measuring $10\times10\times40$ cm, subjected to uniaxial compression with strain increments of $6\times10^{-5}$, see \Cref{fig:sim_wave_validation}(a). 
The specimen dimensions ensure that the ultrasonic transducers are placed at least three wavelengths away from the source, minimizing near-field effects and boundary interference. 
The specimen’s high height-to-width ratio reduces frictional influences, allowing the mechanical behavior to closely represent that of real concrete.
The ultrasonic excitation uses a moment tensor source with two non-zero components (in a 2D model configuration) and a central frequency of 65 kHz, which approximates the actual transducer emission. 
Data acquisition employs 27 receivers arranged across three parallel planes located 27.5, 30.0, and 32.5 cm from the source, with a spacing of 2.5 cm between receivers on each plane, as depicted in \Cref{fig:sim_wave_validation}(a). 
%
Ultrasonic wave propagation in three dimensions is simulated using a rotated staggered-grid finite-difference method.
The simulation runs for 40,000 time steps, covering a total duration of 2 milliseconds. 
A spatial grid resolution of 0.5 mm and a temporal step size of $5\times10^{-8}$ seconds are employed. 
To account for the attenuation effects commonly observed in ultrasonic experiments, a viscoelastic model is incorporated, which involves four additional parameters as detailed in \citep{Saenger_et_al:05}.  
To emulate reflective boundary conditions, two vacuum layers are added around all surfaces of the concrete model. 
As demonstrated in \Cref{fig:sim_wave_validation}(b), the simulated signal exhibit strong qualitative agreement with experimental data, especially in terms of the amplitude decay over time.

\subsubsection{CWI Analysis}
%
Prior to dataset generation for the classification task, two essential preparatory steps are undertaken. 
First, ultrasonic wave propagation simulations in concrete are validated by comparing CWI-derived metrics from simulations with experimental results. This ensures that the simulated signals accurately capture the waveform changes associated with damage progression. This step ensures that the simulated signals reliably capture the evolution of waveforms associated with damage progression, as observed in the experimental data.
The second entails the interpretation of CWI-derived metrics to classify potential damage states in concrete subjected to uniaxial compressive loading.

\begin{figure}[!t]
	\centering
	\includegraphics[scale=1.0]{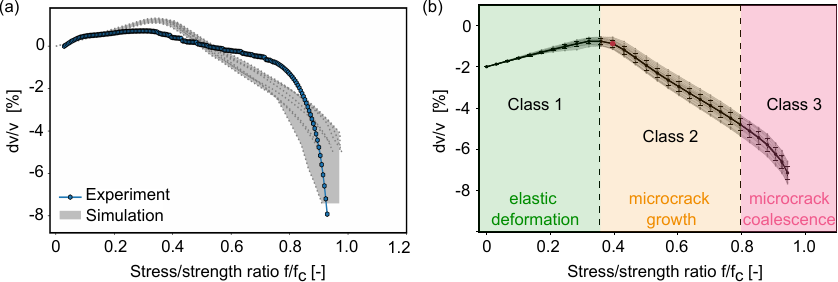}
	\caption{CWI analysis in concrete specimen under compression. (a) Validation of $dv/v$ vs. applied stress at 0.3 ms time window; (b) Definition of damage classes based on CWI analysis of 300 wave simulations.}
	\label{fig:sim_cwi_validation}
\end{figure}

For the validation step, the relative velocity change ($dv/v$) at a specific time window from 0.2 ms to 0.5 ms is plotted as a function of applied stress, as illustrated in \Cref{fig:sim_cwi_validation}(a).
Both simulation and experimental data indicate a peak in $dv/v$ occurring at approximately 35\% of the peak stress. 
However, the numerical model does not fully capture the gradual decline in $dv/v$ observed during the micro-cracking regime, nor the pronounced reduction evident beyond 80\% of the concrete strength. 
These discrepancies may be attributed to several model limitations, including the absence of stable crack propagation mechanisms at interfacial transition zones, an idealized representation of microcrack distributions, and simplified assumptions regarding wave source characteristics and signal recording.
Despite these simplifications, the numerical results exhibit a generally acceptable level of agreement with experimental observations, particularly in terms of capturing the qualitative trend of velocity evolution with increasing stress. 
This level of correspondence supports the use of the simulation framework for subsequent damage classification tasks.

A total of 300 wave propagation simulations were conducted using the described setup, yielding 4500 ultrasonic signals for damage classification via CWI analysis \citep{niederleithinger2018processing}. 
The stretching technique, combined with the auto-CWI method, was applied using a fixed 0–1 ms time window and a cross-correlation threshold of 0.7 to ensure consistent measurements. 
The computed relative velocity variations ($dv/v$) were plotted against normalized stress, revealing three distinct damage stages that define the classification labels as depicted in \Cref{fig:sim_cwi_validation}(b). 
In the initial elastic regime (Class 1), $dv/v$ increases linearly at a rate of 0.084\% per MPa, peaking at 1.235\%, which reflects elastic stiffening due to crack closure and pore compaction. 
In the subsequent phase (Class 2), spanning approximately 35–80\% of the peak load, $dv/v$ begins to decline gradually at –0.112\% per MPa, indicating the development of diffuse microcracking that offsets any local stiffening effects. 
In the final phase (Class 3), beyond 80\% of peak load, $dv/v$ decreases sharply at –0.331\% per MPa, marking the transition to localized damage and rapid stiffness degradation. 
At 42.1\% of peak load, the cross-correlation coefficient drops below 0.7, triggering a reference signal update. 
Compared to experimental data, the simulated onset of damage in Class 2 occurs slightly earlier, potentially due to the absence of stable crack propagation in the Interfacial Transition Zones (ITZs). 
Across all sensor locations, moderate variability in $dv/v$ response is observed, with the standard deviation increasing to 0.34\% at higher load levels, capturing the inherent heterogeneity of damage evolution in concrete. Finally, the simulated dataset $\mathcal{D}_s$ consists of 4500 samples in total, including 1740 samples of Class 1, 1680 samples of Class 2, and 1080 samples of Class 3.

\section{Results and Discussion}
\label{sec:Result}
\subsection{Performance Comparison: Proposed Method vs. Plain 1D CNN}
To better evaluate the performance improvement achieved by the proposed method in concrete damage classification, a plain 1D CNN was implemented as the baseline for comparison. The architecture and training configuration of the plain 1D CNN are kept identical to those of the proposed framework, except that both the DA and BYOL modules were completely removed, resulting in a purely supervised model trained solely on simulated data.

\begin{figure}[!t]
\centering
\includegraphics[scale=1.0]{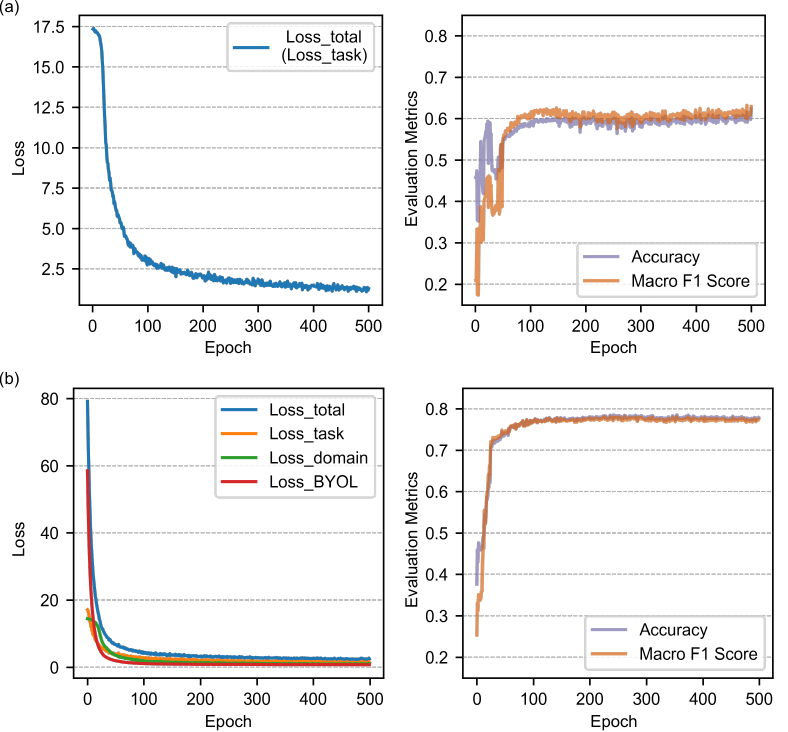}
\caption{Training curves of (a) the plain 1D CNN and (b) the proposed method. Left: convergence of total loss and individual loss components over 500 epochs. Right: corresponding evolution of accuracy and macro F1 score on the experimental data.}
\label{fig:loss}
\end{figure}

Figure~\ref{fig:loss}(a) and \ref{fig:loss}(b) illustrate the training dynamics of the plain 1D CNN and the proposed method, respectively, in terms of both loss convergence and classification performance on the experimental data. After 500 epochs, the total loss in both models converges, indicating successful optimization. For the proposed method, all components of the total loss stabilize over time, demonstrating that the joint training process of all modules remains stable and balanced throughout. Moreover, compared to the plain 1D CNN, the proposed method achieves not only higher accuracy and macro F1 scores on the experimental data, but also markedly reduced fluctuations throughout training, showing improved convergence behavior and greater training stability under domain shift.

\begin{figure}[!t]
\centering
\includegraphics[scale=1.0]{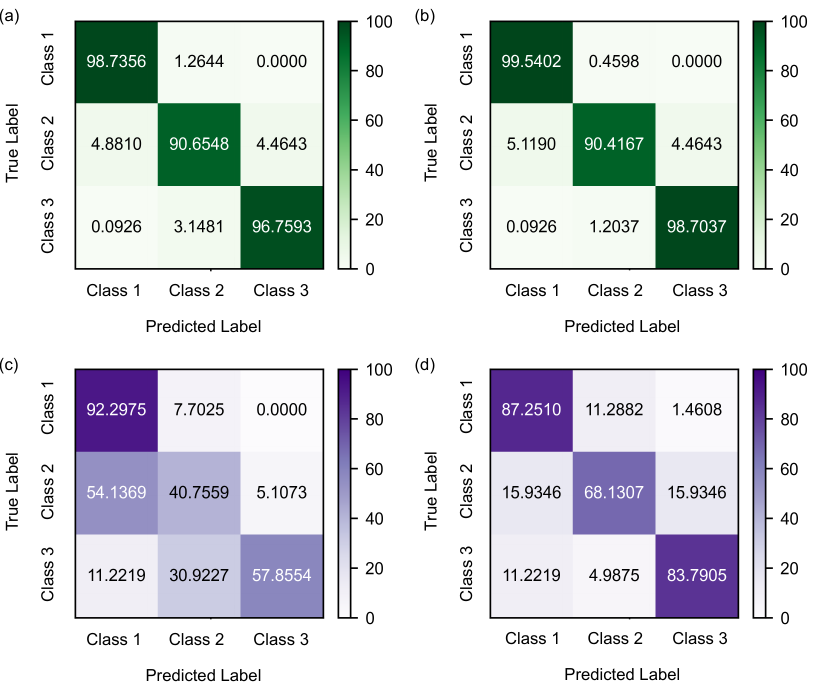}
\caption{Confusion matrices of the plain 1D CNN and the proposed method on simulated and experimental datasets. (a) Plain 1D CNN on simulated data; (b) Proposed method on simulated data; (c) Plain 1D CNN on experimental data; (d) Proposed method on experimental data.}
\label{fig:confusion}
\end{figure}

Figures~\ref{fig:confusion}(a)-(d) compare the confusion matrices of the plain 1D CNN and the proposed method on both simulated and experimental datasets. On the simulated data (Figs.~\ref{fig:confusion}(a) and (b), both models achieve high classification accuracy across all classes, indicating that they can effectively learn from the source domain. However, a notable performance gap is observed when evaluated on the experimental data. The baseline model (Fig.\ref{fig:confusion}(c)) shows excellent performance in identifying Class 1 (92.30\%) but fails to generalize to Classes 2 and 3. In particular, 54.14\% of Class 2 samples are misclassified as Class 1, and 30.92\% of Class 3 samples are misclassified as Class 2, indicating poor cross-domain generalization and insufficient inter-class separability within the target domain. In contrast, the proposed method (Fig.~\ref{fig:confusion}(d)) significantly reduces misclassification across all classes. It achieves well-balanced predictions with class-wise accuracies of 87.38\%, 68.13\%, and 83.79\% for Classes 1, 2, and 3, respectively. 

Quantitatively, the plain 1D CNN achieves an overall accuracy of 0.6240 and a macro F1 score of 0.6294 on the experimental data. By comparison, the proposed method improves these metrics to 0.7803 and 0.7773, respectively, representing a substantial enhancement in classification performance. These results emphasize the effectiveness of the proposed domain adaptation framework in mitigating domain shift and achieving label-free concrete damage classification under real-world experimental conditions.

In terms of computational cost, the training time for the plain 1D CNN is approximately 1 minute and 10 seconds, while the proposed method requires 3 minutes and 9 seconds. Despite the additional neural network modules involved, this moderate increase in computational overhead is highly acceptable given the notable performance improvements. The proposed approach thus provides a practical and efficient solution for transferring knowledge from the simulation domain to the experiment domain, enabling reliable concrete damage identification based on coda waves.

\subsection{Investigation on the Robustness of the Proposed Method}
\begin{figure}[!t]
\centering
\includegraphics[scale=1.0]{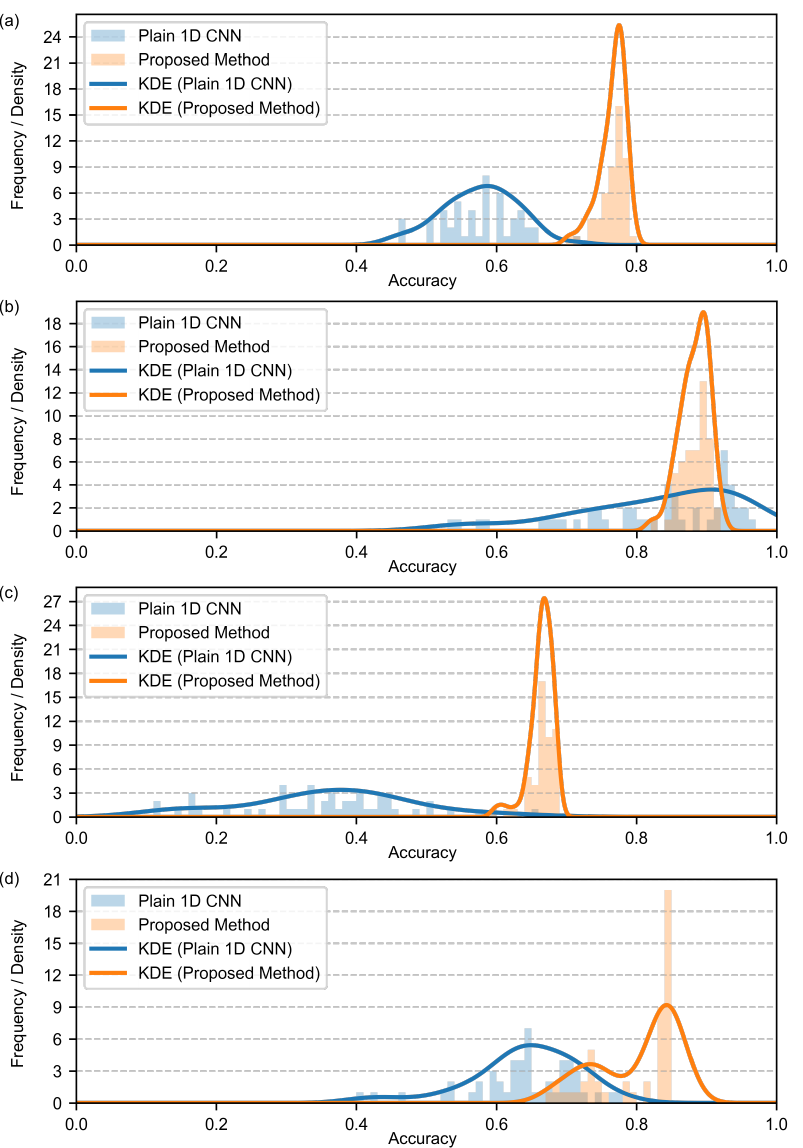}
\caption{Comparison of classification robustness between the plain 1D CNN and the proposed method using histogram and kernel density estimation (KDE). Subfigures (a)–(d) illustrate the distribution of overall accuracy and class-wise accuracies for Classes 1, 2, and 3, respectively.}
\label{fig:histogram}
\end{figure}

This section explores the robustness of the proposed method, both the plain 1D CNN and our method were independently trained and tested 50 times using different random seeds under identical experimental settings. Figure~\ref{fig:histogram} presents the resulting accuracy distributions, which are visualized through histograms and kernel density estimate (KDE) plots. 

As shown in Figure~\ref{fig:histogram}(a), the proposed method exhibits a highly concentrated accuracy distribution centered around 0.78, with low variance and a sharp unimodal KDE curve. In contrast, the plain 1D CNN shows a much broader distribution centered near 0.58, with a heavier left tail and more low-accuracy outliers. This contrast highlights the superior consistency and reliability in overall classification performance of the proposed method.

Figure~\ref{fig:histogram}(b) presents distinct distribution characteristics for Class 1 between the two models. While both models achieve relatively high accuracies, the proposed method demonstrates much tighter distribution, with its KDE sharply peaking around 0.88. In comparison, the baseline model shows a flatter and more dispersed distribution, suggesting greater sensitivity to variations in initialization or optimization dynamics across runs.

In Figure~\ref{fig:histogram}(c), which corresponds to Class 2, the performance gap becomes even more evident. The plain 1D CNN produces a wide and left-skewed distribution, with many runs falling below 0.4 in accuracy. On the other hand, the proposed method maintains a narrow distribution centered around 0.67, demonstrating a substantially improved ability to extract domain-invariant features for this particularly challenging Class.

Figure~\ref{fig:histogram}(d) presents the results for Class 3. The baseline model yields a distribution with moderate variance, reflecting fluctuations in performance across runs. In contrast, the proposed method exhibits a sharper peak near 0.84, with the majority of runs concentrated around this high-accuracy region. This indicates that the proposed method not only achieves better average performance but also delivers more stable and reliable predictions for Class 3.

In summary, the proposed method outperforms the plain 1D CNN in terms of both accuracy and robustness. Across all Classes, it not only improves average classification accuracy but also significantly reduces inter-run variability. Such robustness is particularly important in real-world SHM applications, where accurate and stable predictions are critical for structural integrity assessment and lifecycle management.

\subsection{Ablation Study}
\begin{figure}[!t]
\centering
\includegraphics[scale=1.0]{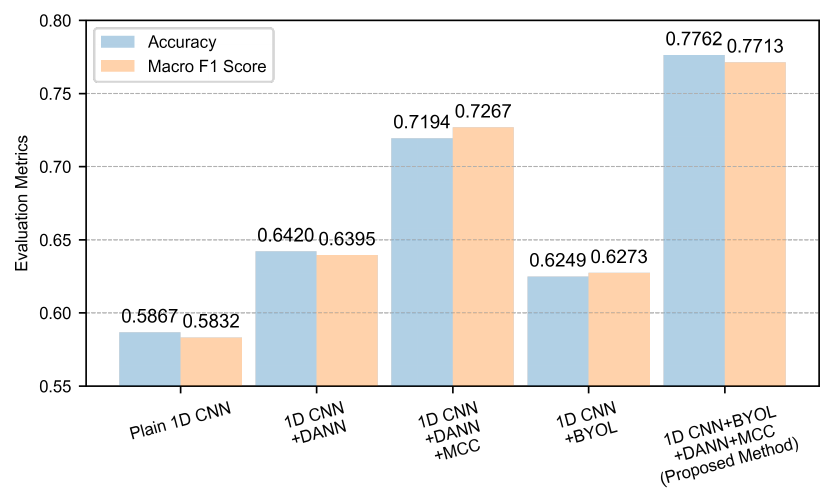}
\caption{Results of the ablation study comparing the impact of different components in the proposed framework. The full model (rightmost group) combines Domain-Adversarial Neural Network (DANN), Minimum Class Confusion (MCC), and self-supervised learning (i.e., BYOL), and outperforms all other configurations in both accuracy and macro F1 score.}
\label{fig:ablation}
\end{figure}

In this section, an ablation study is carried out to evaluate the individual and joint contributions of its core components.  Five model configurations are considered: (1) plain 1D CNN, (2) 1D CNN with DANN only, (3) 1D CNN with DANN and MCC, (4) 1D CNN with self-supervised learning via BYOL, and (5) the complete model combining DANN, MCC, and BYOL, referred to as the proposed method. We run each model for 10 times and report the average results. The performance comparison in terms of accuracy and macro F1 score is illustrated in Fig.~\ref{fig:ablation}.

The plain 1D CNN yields the lowest performance, with an accuracy of 0.5867 and a macro F1 score of 0.5832, reflecting its limited ability to generalize from simulated to experimental data. Introducing DANN without MCC leads to moderate improvement (accuracy: 0.6420, F1: 0.6395), indicating that domain adversarial training partially mitigates distribution shift but is still suboptimal when class confusion remains unaddressed.

When MCC is incorporated into the domain adaptation process, the performance improves significantly (accuracy: 0.7194, F1: 0.7267), confirming the benefit of reducing inter-class confusion in enhancing cross-domain classification. On the other hand, implementing BYOL alone also leads improvements over the baseline (accuracy: 0.6249, F1: 0.6273), suggesting that self-supervised learning facilitates the extraction of domain-transferable features from unlabeled experimental data.

Finally, the proposed method, which integrates DANN, MCC, and BYOL, achieves the highest overall performance (accuracy: 0.7762, macro F1: 0.7713). This demonstrate that the three components work synergistically to improve the generalization under domain shift. All these results validate the soundness of the framework design and emphasize the importance of jointly addressing feature alignment, class confusion reduction, and self supervised learning for accurate and reliable concrete damage classification in real-world scenarios. 

\subsection{Comparison with Representative Domain Adaptation Methods}
The superiority of the proposed method is evaluated through comparison with six representative DA techniques. All models adopt the same base architecture and are trained following the settings recommended in their original publications. For each model, results are averaged over ten independent runs to ensure statistical reliability. Specifically, the comparative models used are listed as follows:
\begin{enumerate}[label=(\arabic*)]
\item 
JAN \citep{Long_2017_JAN}: JAN (Joint Adaptation Network) alleviates domain shift by matching the joint distributions of multiple domain-specific layers across source and target domains using Joint Maximum Mean Discrepancy (JMMD).
\item 
MCD \citep{Saito_2018_MCD}: MCD (Maximum Classifier Discrepancy) attempts to align source and target distributions by utilizing task-specific decision boundaries: two classifiers are trained to maximize their prediction discrepancy on target samples, while a feature generator learns to minimize this discrepancy in an adversarial manner.
\item  
MDD \citep{Zhang_2019_MDD}: MDD (Margin Disparity Discrepancy) is a theoretically justified measurement that compares source and target distributions with asymmetric margin loss. It can be seamlessly integrated into an adversarial learning algorithm for domain adaptation.
\item
DSAN \citep{Zhu_2021_DSAN}: DSAN (Deep Subdomain Adaptation Network) is a non-adversarial method that enhances DA by aligning subdomain distributions across source and target domains using Local Maximum Mean Discrepancy (LMMD).
\item 
MMSD \citep{Qian_2023_MMSD}: MMSD (Maximum Mean Square Discrepancy) is a novel discrepancy representation metric that extends traditional MMD by using both the mean and variance information of data distributions in reproducing kernel Hilbert space (RKHS). 
\item 
DALN \citep{Chen_2022_DALN}: DALN (Discriminator-free Adversarial Learning Network) reuses the task classifier as a discriminator to achieve explicit domain alignment and category distinguishment through a unified objective. It further introduces Nuclear-norm Wasserstein Discrepancy (NWD) to enhance prediction determinacy and diversity.
\end{enumerate}

\begin{figure}[!t]
\centering
\includegraphics[scale=1.0]{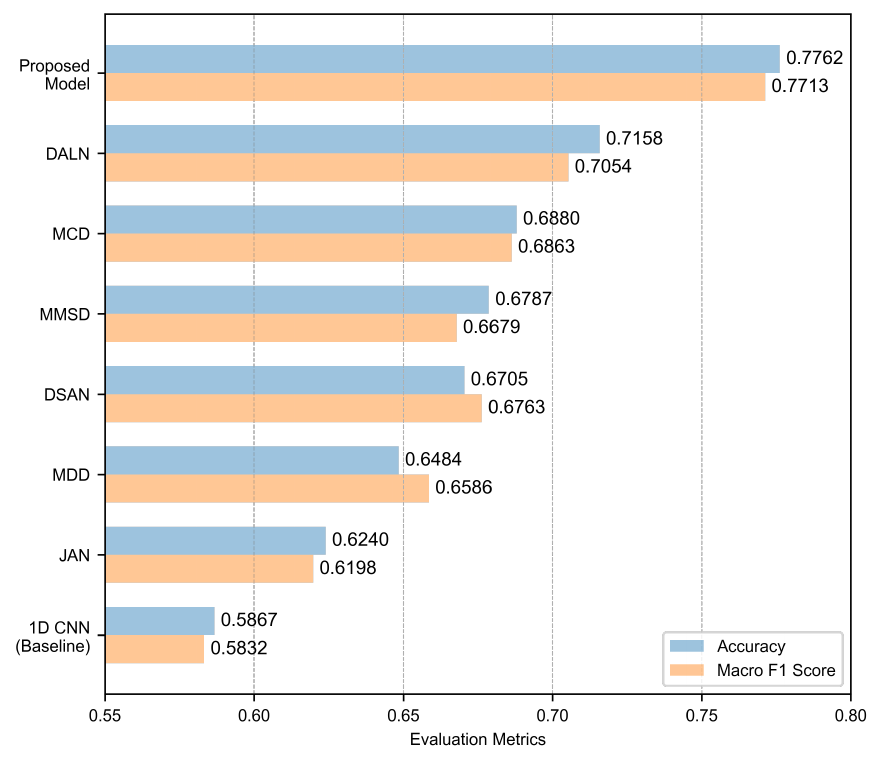}
\caption{Performance comparison between the proposed method and six representative domain adaptation techniques on the experimental dataset.}
\label{fig:model_comparison}
\end{figure}

Figure~\ref{fig:model_comparison} presents the average results of these methods over ten runs in terms of accuracy and macro F1 score. All DA methods achieve noticeable improvements over the plain 1D CNN baseline. Among all the approaches evaluated, the proposed method has the best performance, with an accuracy of 0.7762 and a macro F1 score of 0.7713, which outperforms all other DA models. Notably, DALN also shows competitive results, with both metrics exceeding 0.7, and performs comparably to the 1D CNN+DANN+MCC configuration discussed in the previous ablation study section.

\begin{figure}[!t]
\centering
\includegraphics[scale=1.0]{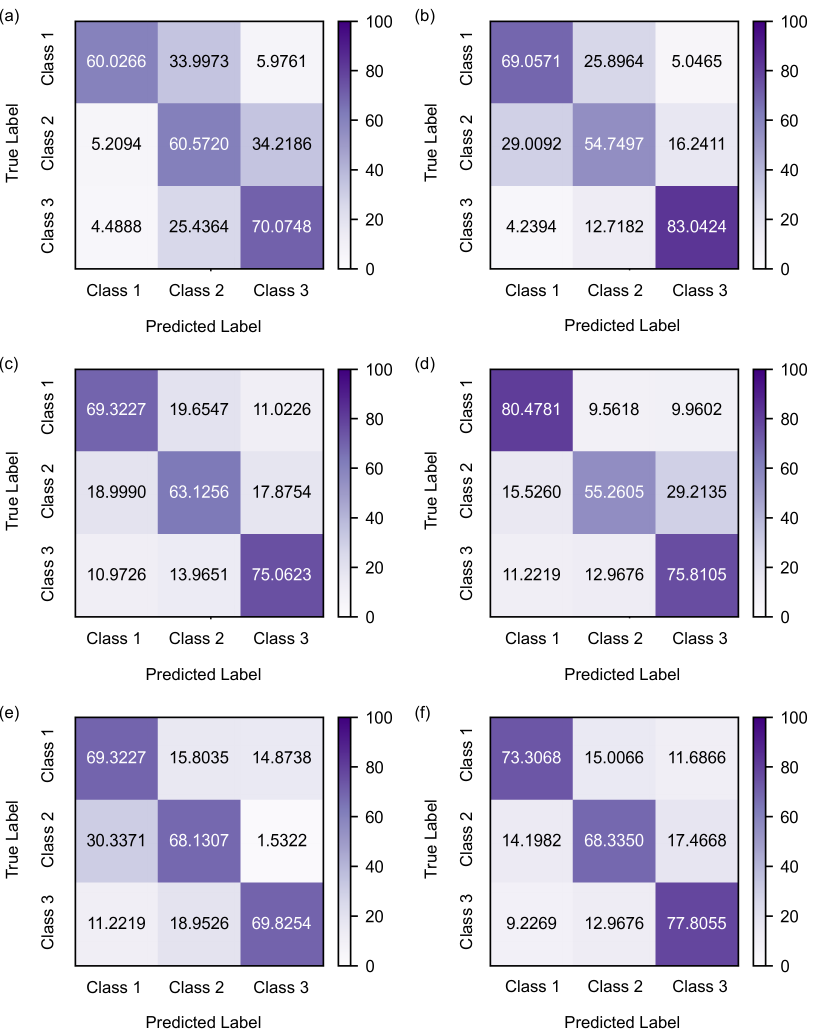}
\caption{Confusion matrices of six representative domain adaptation methods on the experimental dataset: (a) JAN, (b) MDD, (c) DSAN, (d) MMSD, (e) MCD, and (f) DALN.}
\label{fig:confu_compare}
\end{figure}

Figure~\ref{fig:confu_compare} provides an overview of the confusion matrices corresponding to the six evaluated DA methods. In all cases, the diagonal elements dominate the matrices, indicating that each model achieves reasonable prediction accuracy across the three classes. Compared to the plain 1D CNN baseline (see Fig.~\ref{fig:confusion}(c)), all DA methods achieve moderate gains in classification performance.
Among the six methods, DALN and MCD exhibit relatively higher and more balanced accuracy across all classes. In particular, both methods exceed 68\% accuracy for Class 2, the most challenging due to its intermediate damage state, suggesting a stronger generalization to the target domain. Nevertheless, none of the methods surpass the proposed approach (see Fig.~\ref{fig:confusion}(d)), which achieves the highest overall accuracy and shows the most distinct separability between classes.

\section{Conclusions}
\label{sec:Conc}
This work proposes a self-supervised domain adaptation framework for robust concrete damage classification. A virtual testing platform is firstly developed, which combines multiscale modeling of concrete damage with ultrasonic wave propagation simulations to generate abundant labeled synthetic data.
The proposed approach bridges the domain gap between simulated and experimental data by integrating domain adversarial training, minimum class confusion, and the BYOL self-supervised learning paradigm, effectively transferring knowledge from the labeled simulated domain to the unlabeled experimental domain. Its key advantages are summarized as follows:
\begin{enumerate}[label=(\arabic*)]
\item
Accurate: The proposed method achieves an accuracy of 0.7762 and a macro F1 score of 0.7713, representing a significant improvement over the baseline plain 1D CNN (accuracy: 0.5867; macro F1 score: 0.5832).
\item
Robust: Our method shows strong robustness across repeated trials. The resulting accuracy distributions are sharply peaked and narrow, as shown in the histogram and KDE plots, indicating consistent performance and stable convergence behavior under domain shift.
\item 
Fast: The framework has high computational efficiency. The training time increases by only about two minutes compared to the plain 1D CNN baseline. It is trained in an end-to-end manner without the need for separate offline pre-training or multi-stage optimization, making it well-suited for practical SHM applications.
\item
Superior: Comparative evaluation against six representative domain adaptation methods (JAN, MDD, DSAN, MMSD, MCD, and DALN) highlights the superior generalization performance of the proposed framework under domain shift.
\end{enumerate}

In future work, additional physical features or multimodal sensing data can be incorporated to further enrich the model input and enhance classification performance. Another important direction is to address class imbalance in real-world data, which may affect the model's generalization under practical scenarios. Moreover, since the current dataset mainly focuses on compressive damage in concrete, extending the framework to include tensile damage scenarios will be an important step toward broader applicability in structural health monitoring. 

While this study is conducted under controlled laboratory conditions, real-world structures often involve more complex boundary conditions and long-term degradation mechanisms. Therefore, adapting the proposed framework for in-situ monitoring applications can also be an important future direction. In this context, knowledge gained from specimen-level damage characterization is expected to serve as a foundation for developing advanced strategies for structural-level damage identification, thereby enabling more effective and reliable assessment of concrete structures in practice.

\section{Acknowledgment}
The first author acknowledges the support from the China Scholarship Council (CSC) under grant 202006260038, the Bridging Scholarship from the Research Department Subsurface Modeling \& Engineering (RDSME) of Ruhr University Bochum, and the Scholarship of Wilhelm and Günter Esser Foundation from the Research School of Ruhr University Bochum. The authors gratefully acknowledge the exchange and numerous discussion with the Chair of Materials Science and Testing at Technical University Munich, in particular with Prof. Christoph Gehlen, which were helpful for the interpretation of experimental data. The authors would also like to thank the German Research Foundation (DFG) for their financial support in the framework of Subprojects RUB1, TUM1 of the Research Unit FOR 2825 (project number 398216472).



\clearpage
\bibliographystyle{apalike}
\bibliography{export.bib}







\end{document}